\documentclass{article}
\usepackage{quel-art}
\usepackage{graphicx}
\usepackage{amsmath}
\usepackage{amssymb}
\usepackage{bm}
\usepackage{epstopdf}

\volnumber{20} \issuenumber{1} \edyear{2004}                             
\frompage{000} \topage{000}                                              
\recrevdate{1 January 2004}                                              
\def\rightmark{Complex chaos in conditional qubit dynamics and purification}
\def\leftmark{T. Kiss, I. Jex, G. Alber and S. Vym\v etal}
 
\title{Complex chaos in conditional qubit dynamics and purification protocols} 
\authors{ 
{T. Kiss$^{(1)}$, I. Jex$^{(2)}$, G. Alber$^{(3)}$ and S. Vym\v etal$^{(2)}$
\index{Kiss, T.} 
\index{Jex, I.} 
\index{Alber, G.}
\index{Vym\v etal, S.}
}\\[2.812mm]
{\normalsize
\hspace*{-8pt}$^{(1)}$~Research Institute for Solid State Physics and
  Optics, H-1525 Budapest, P. O. Box 49, Hungary\\ %
\hspace*{-8pt}$^{(2)}$~Department
  of Physics, FJFI \v CVUT, B\v rehov\'a 7, 115 19 Praha 1 - Star\'e
  M\v{e}sto, Czech Republic\\ 
\hspace*{-8pt}$^{(3)}$~Institut f\"ur Angewandte
  Physik, Technische Universit\"at Darmstadt, D-64289 Darmstadt,
  Germany 
}}
 
\abstract{Selection of an ensemble of equally prepared quantum
  systems, based on measurements on it, is a basic step in quantum
  state purification. For an ensemble of single qubits, iterative
  application of selective dynamics has been shown to lead to complex
  chaos, which is a novel form of quantum chaos with true sensitivity
  to the initial conditions. The Julia set of initial valuse with no convergence shows a complicated structre on the complex plane. The shape of the Julia set varies with the parameter of the dynamics. We present here results for the two qubit case demonstrating how
  a purification process can be destroyed with chaotic oscillations.  }
  

 
\makeindex
\begin{document}
 
\maketitle

\section{Introduction}\label{intro}

Quantum state purification is a process where entanglement is enhanced
in an ensemble of quantum states, by making measurements on part of
the ensemble and using the gained information to select part of the
ensemble \cite{purification}. The complete evolution of the remaining
subensemble is not unitary anymore, since the measurement process
extracts information from the system on one hand, and this information
is fed back by the conditional selection itself. In this way,
nonlinear transformations can be realized \cite{Gisin98}.
 
Chaos in dynamical systems has a long history in physics
\cite{Poincare1892}.  Exponential sensitivity to initial conditions is
the usual definition for chaos in classical physics.  In quantum
mechanics, however, unitary evolution of a closed system prevents
initially close states to separate exponentially fast, nevertheless
quantum chaos is an important field dealing with quantized
counterparts of classically chaotic systems
\cite{LesHouches91}. 

Occurrence of true chaos should be possible in a
model describing quantum-classical correspondence, for example via
continuous measurements, as confirmed by recent numerical results
\cite{quantum-classical}.  Moreover, it is possible to find a regime
where the continuously measured quantum system is far from the
classical limit and still it exhibits true chaos with a positive
Lyapunov exponent, which was calculated by numerical simulations
\cite{HJS}. Continuous measurements are used to model the environmental decoherence and the measurement result are not fed back to the system, resulting in stochastic dynamics.

Recently, we have proven \cite{KJAV} that true chaos occurs in the
conditional dynamics of qubits, in a similar arrangement proposed for
quantum state purification \cite{Gisin98}.  We could calculate the
positive Lyapunov exponent analytically for a simple case. The more
general dynamics of a single qubit is described by a complex,
nonlinear map, in this way directly realizing complex chaos
\cite{KJAV}.  This situation is different compared to classical chatotic dynamics, in the sense that classical chatotic maps are acting on the 
phase space with real numbers. The complex nonlinear map, behind complex chaos, is acting on the complex space representing the Hilbert space of the system. We are not aware of any other physical system directly realizing complex chaotic maps.

Conditional dynamics by measurement selection as applied
in purification can be considered as feeding back results of strong
(von Neumann) measurements with the selection as opposed to the idea
of feeding back the results of weak measurements \cite{Lloyd2000}. The
latter scheme has also been suggested as a candidate for a truly
chaotic quantum system, although no proof was provided.

In the present paper we focus on implications of complex chaos for the
purification procedure.  The structure of the paper is the
following. First, we shortly review how complex chaos occurs in
iterated deterministic quantum maps. For a one qubit system we
illustrate the behaviour of the system by showing the nontrivial Julia
sets of non-converging initial states. In the next section, we
consider the two qubit system as an example of quantum state
purification and finally we conclude.

\section{Iterated deterministic quantum maps}\label{techno}  

The quantum XOR gate is defined as the following operation
\begin{equation}
\label{ }
XOR_{12}|i\rangle_{1} |j \rangle_{2 }= |i \rangle_{1} |i-j \rangle_{2}
\mod(D)\, ,
\end{equation}
where $D$ is the dimension of the Hilbert space for both subsystems.
The XOR gate is the key operation for deterministic quantum state
purification protocols \cite{Gisin98} together with a measurement. For
a nonlinear transformation one needs two identical copies of the
system. The application of the XOR gate followed by the measurement of
the second system in the $|0\rangle$ state leaves the first system in
a state defined by the following nonlinear transformation
\begin{equation}
\rho ^\prime = {\cal S}\rho\, , \quad \rho_{ij} \xrightarrow{\cal S} N
\rho_{ij}^2\,
\label{maprho}
\end{equation}
with the renormalization factor $N=1/\sum\rho_{ii}^2$.  The matrix
element squaring is defined with respect to a prescribed orthonormal
basis $\{|i\rangle\}$.  By repeating the transformation ${\cal S}$ we
expect that smaller diagonal matrix elements will tend to zero and the
largest one survives, converging to unity. In other words, pure states
can be stable fixed points of the map, leading to purification of the
state.

Let us consider now the $D=2$ case, when each system is a qubit. The
most general iterative deterministic dynamics based on matrix element
squaring is achieved if we allow for an arbitrary unitary
transformation in each step
\begin{equation}
\label{rotation1}
{\cal R} \rho = U \rho U^{\dagger} \, ,
\end{equation}
with 
\begin{equation}
U = \left(
\begin{array}{lll} 
&\cos x &  \sin x~e^{i\phi}\\
-&\sin x~e^{-i\phi} & \cos x
\end{array}
\right) \, ,
\label{rotation2}
\end{equation}
in the prescribed basis.  In this way one step of the dynamics reads
\begin{equation}
\rho^{\prime} = {\cal F} \rho = {\cal R S} \rho \, ,
\end{equation}
and repeating the transformation ${\cal F}$ leads to the discrete
conditional time-evolution we are interested in.

The description of the dynamics simplifies considerably if we take an
initial pure state of the qubit. Let us introduce the following
notation
\begin{equation}
|\psi\rangle = N (z |0\rangle + |1\rangle)\, ,
\end{equation}
where the state is normalized by $N=(1+|z|^{2})^{-1/2}$. Thus we haved
represented the qubit Hilbert space on the complex plane, where the
parameter $z$ describes the state of the qubit. The transformation
$\cal F$ maps the pure state onto a pure state and transforms $z$ as
\begin{equation}
\label{mapz}
z \mapsto F_{p}(z)= \frac{z^{2}+p}{1-p^{*}z^{2}}\, ,
\end{equation}
where $p=\tan x e^{i \phi}$ and the star denotes complex conjugation.
The conditional dynamics of the qubit are thus governed by $F_{p}(z)$,
which is a non-linear $\mathbb{C}\to\mathbb{C}$ map with one complex
parameter $p$. A considerable difference compared to chaotic systems
in classical physics is that the underlying space is complex,
here. Even the simplest non-linear maps of the complex plane can show
intricate dynamical structure, such as the famous Mandelbrot set.  The
study of the mathematics related to maps in one complex variable has a
long history and an extensive literature, (for a review see
\cite{Milnor}).

The traditional approach to a non-linear map in one complex variable
is to divide the complex plane of the initial values $z_{0}$ into
regular and irregular points forming the Fatou and Julia sets,
respectively. Regular starting points from the Fatou set will converge
to a stable cycle (also elements of the Fatou set) when repeating the
iteration. Initial values included in the Julia set are considered to
be chaotic, leading to irregular oscillations or forming unstable
cycles. Taking into account both the initial condition $z_{0}$ and the
complex parameter $p$ a four dimensional parameter space is
defined. In a simplified situation, when the parameter $p$ is set to
zero, we could define and calculate the positive Lyapunov exponent
\cite{KJAV}.
\begin{figure}
\includegraphics[width=6.6cm]{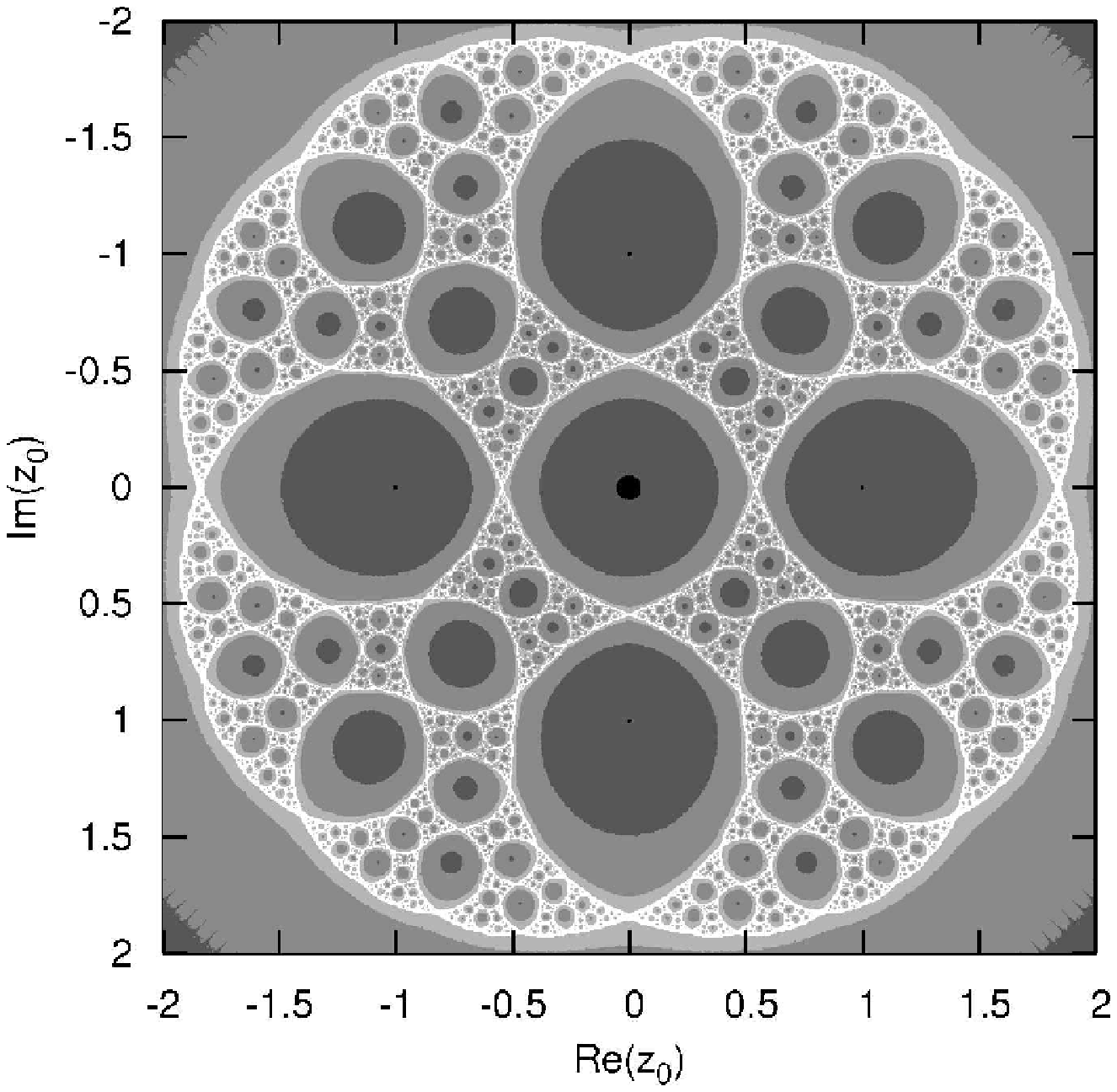}
\includegraphics[width=6.8cm]{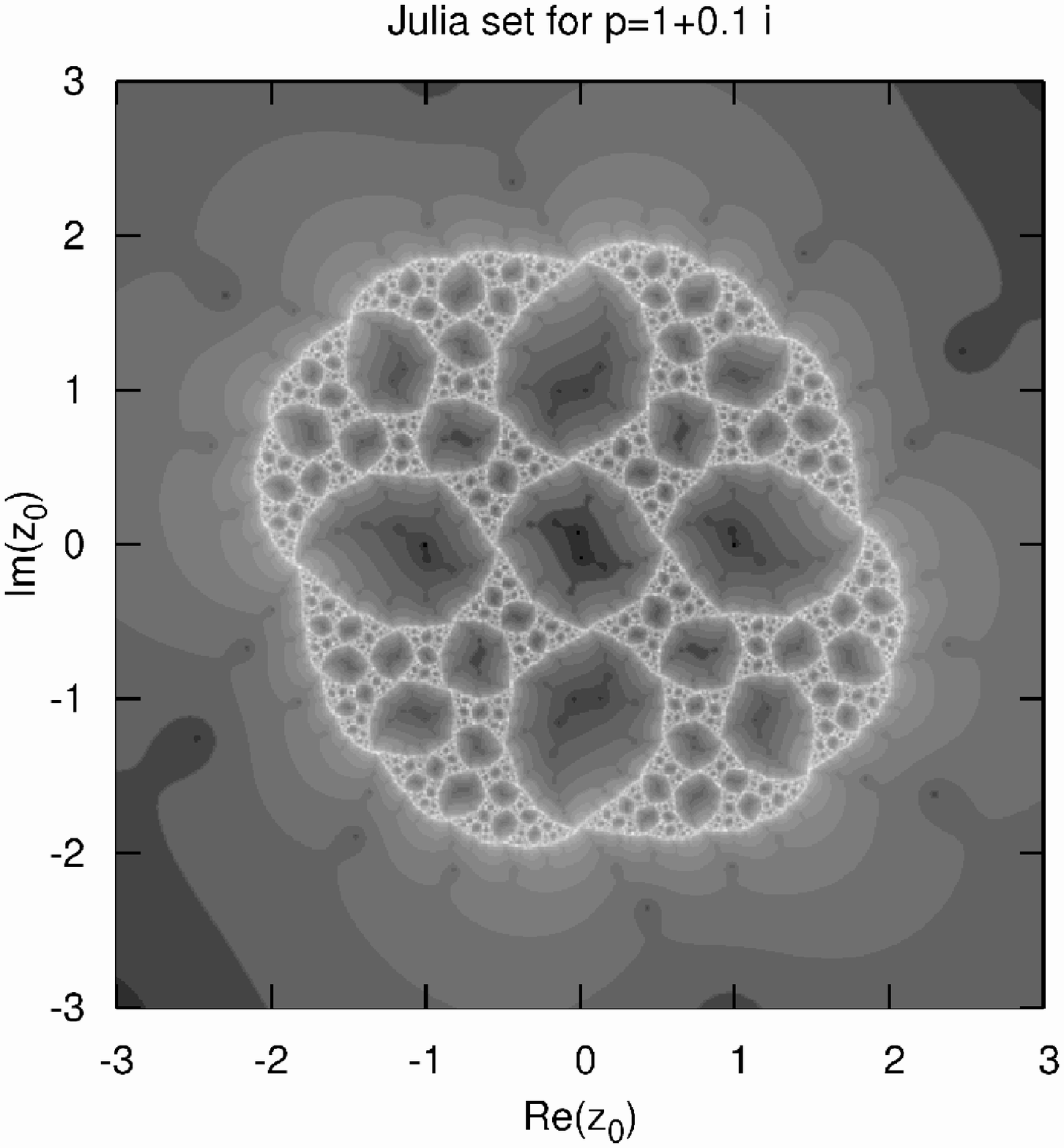}\\
\includegraphics[width=6.5cm]{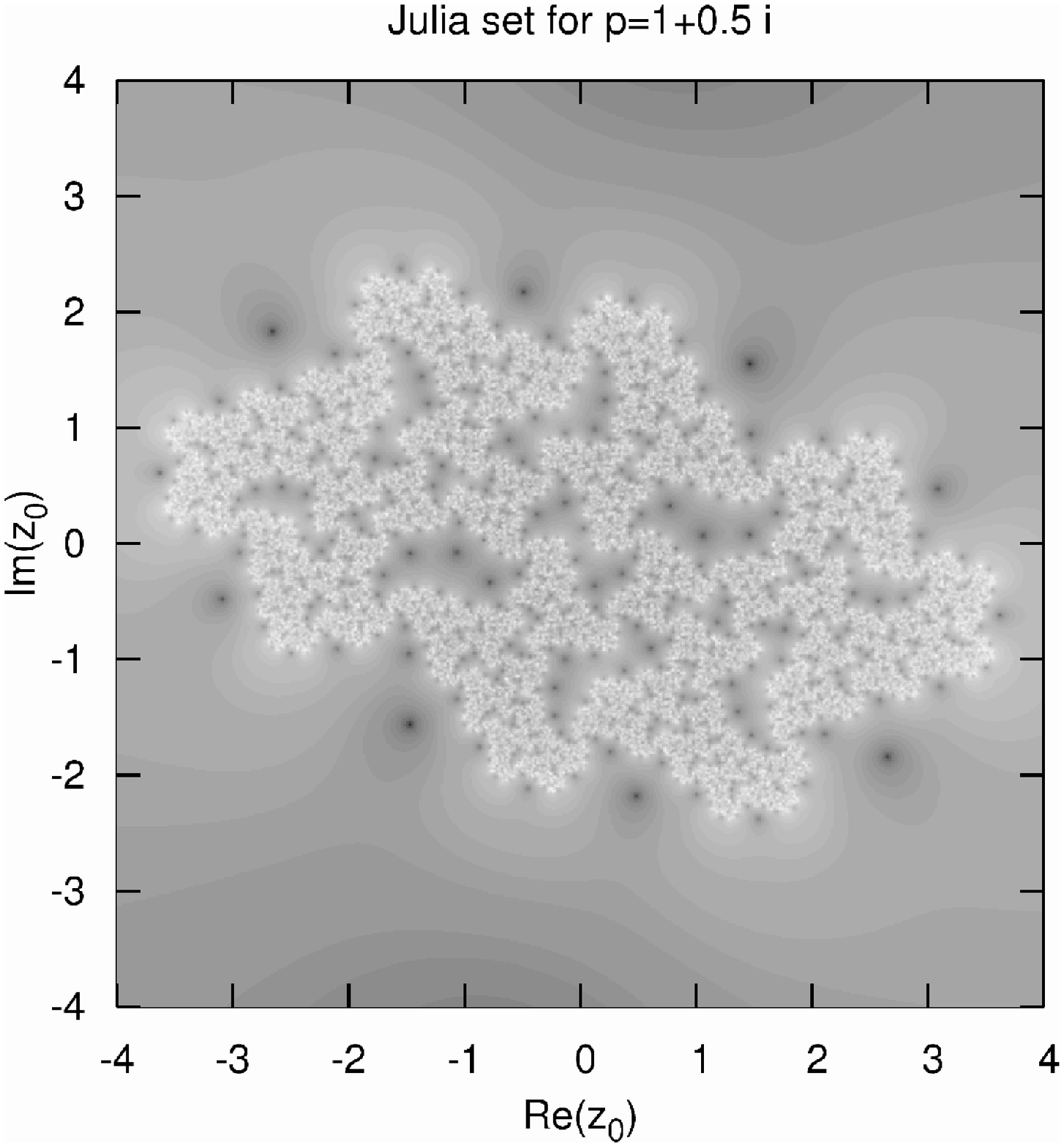}
\includegraphics[width=6.7cm]{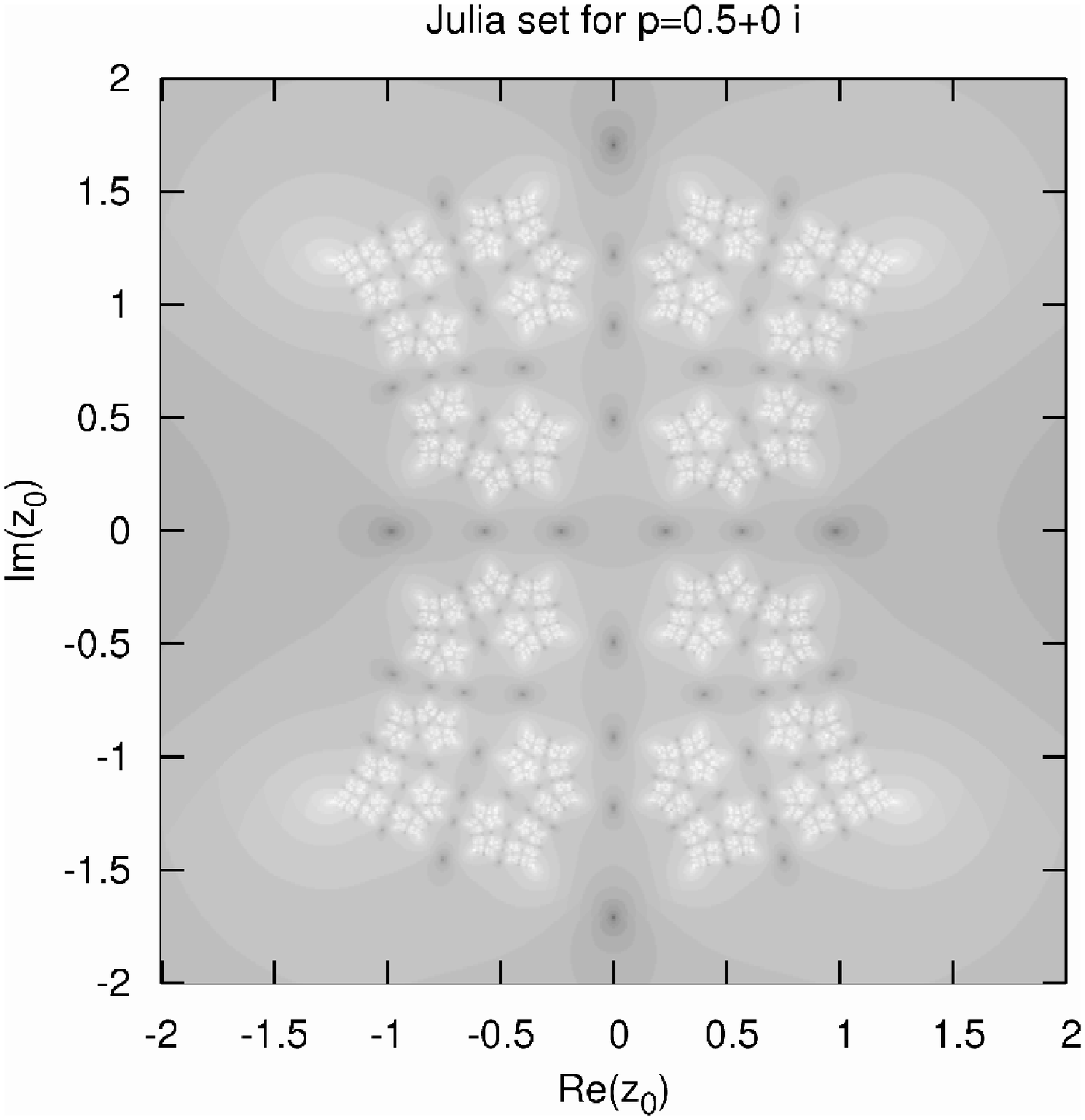}
\caption{ Julia sets for the non-linear map (\ref{mapz}). The
  parameter is set to $p=1$ on the upper left, $p=1+0.1 I$ on the
  upper right, $p=1+0.5 I$ on the lower left and $p=0.5$ for the lower
  right picture. Grayscale indicates how fast the map converges to the
  stable cycle $\{-1,\infty\}$ (dark -- fast, grey -- slow
  convergence, white -- no convergence).}
\label{Fig-julia}
\end{figure}

The full dynamics induced by Eq. (\ref{mapz}) take place on the
Riemann sphere $\hat{\mathbb{C}}$ consisting of $\mathbb{C}$ together
with the point at infinity. The physical meaning of the points $0$ and
$\infty$ for $z$ are the two basis states of the qubit, $|1\rangle$
and $|0\rangle$, respectively. The map $F_{p}(z)$ is a rational
function of degree two.  The Julia sets corresponding to various $p$
values can possess highly non-trivial structures.  Physically
speaking, the parameter $p$ describes the rotation of the qubit state
$|\psi \rangle$. Setting the parameter to $p=1$ corresponds to a
rotation of $\pi/4$ that transforms, for example, the basis states
into their equal superpositions.  This is a symmetric situation with
respect to the basis states. The orbit of one critical point,
$z_{c2}=\infty$, is part of the attractive cycle $\{-1,\infty\}$. The
other critical point $z_{c1}=0$ follows the orbit $0 \mapsto 1 \mapsto
\infty$ and thus lands on the same periodic cycle.  Therefore the only
stable cycle for this map is the fixed point $\{-1,\infty\}$. This
also proves that the map is hyperbolic \cite{Milnor}.  In general,
numerical calculation of the Julia set for quadratic rational maps is
difficult. There are no generic algorithms to compute it. Here we can
simply apply the criterion of convergence to the stable cycle.  In
Fig.(\ref{Fig-julia}) we show the Julia set for $p=1$ as well as for
$p=1+0.1 I$, $p=1+0.5 I$ and $p=0.5$.  The first 3 cases are similar
to each other, there is one stable periodic cycle with length 2. The
sets look like gradual distortion of the initial set. In the last case
we have also only one stable cycle, but its length is 1. The structure
of the set is qualitatively different from the previous cases. These
figures illustrate well the sensitivity for initial conditions, a
small change of the initial state can change the convergence
properties of the dynamics.

\section{Purification protocol}\label{others}

Purification of quantum states usually aims at increasing entanglement
within a system. A system consisting of two qubits was considered with
conditional dynamics for purification \cite{Gisin98}.  One would
expect that sensitivity found in the one qubit case will also occur
here and affect purification. Let us consider the following iteration
acting on two-qubit states
\begin{equation}
\label{twobitmap}
\rho^{\prime} = {\cal F} \rho = {\cal R}_1 {\cal R}_2 {\cal S} \rho \, .
\end{equation}
Here ${\cal S}$ is the element squaring defined in
Eq. (\ref{maprho}) with the index $i$ running from 1 to 4 through the
elements of the product basis of the qubits $\{|j\rangle|k\rangle\}$,
$(j,k=0,1)$ and the rotation ${\cal R}_m$ acts on the $m$th qubit,
with parameters $x_m \, , \phi_m$ as defined in
Eqs. (\ref{rotation1},\ref{rotation2}).  Now, the parameters of the
two (local) complex rotations span $\mathbb{C}^{2}$, and the initial
state can be any valid two-qubit density operator. Obviously, this is
an even much larger parameter space to explore, which includes the
one-qubit pure states as a special case.

\begin{figure}
\begin{center}
\includegraphics[width=11cm]{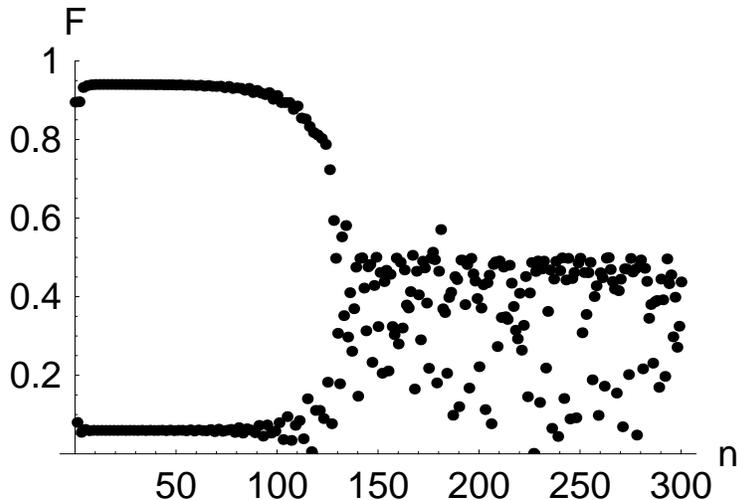}
\end{center}
\caption{Fidelity of purification for a pair of qubits by the
  non-linear map (\ref{twobitmap}), with rotation angles
  $x_{1}=x_{2}=0.293 \pi$, $\phi_{1}=\phi_{2}=\pi/4$ and initial state
  defined in Eq. (\ref{initial}) with fidelity
  ($F_{0}=0.895$). Irregular dynamics set in after a rather long
  period-two transient.  }
\label{Fig-twobits}
\end{figure}

In order to demonstrate how a purification process may be affected by
the chaotic nature of this non-linear map, we choose a target state
$|\psi_{target} \rangle = \frac{1}{\sqrt{2}} (|0 1 \rangle + | 1 0
\rangle)$. This state is part of a second order fixed point of the map
when all four rotation angles equal to $\pi/4$. To characterize
purification we apply the fidelity defined by the overlap with the
target state $F = Tr (\rho \rho_{target})$. Let us start from a mixed
initial state slightly different from $|\psi_{target} \rangle$,
described by the density matrix
\begin{equation}
\rho_{0} = \left(
  \begin{array}{cccc}
    0.1 & 0 & 0 & 0 \\ 
    0 & 0.45 & 0.445 & 0 \\ 
    0 & 0.445 & 0.45 & 0 \\ 
    0 & 0 & 0 & 0 \\ 
  \end{array}
\right)\, .
\label{initial}
\end{equation}
During the evolution, the fidelity grows in every second step, tending
to 1, while the other member of the limit cycle is an orthogonal
state. This can be considered as a purification process.  If we now
slightly change the angles of both qubit rotations $x_{1}=x_{2}$, the
oscillations become metastable, and after a finite number of
iterations irregular behavior emerges as shown in
Fig.~\ref{Fig-twobits}.
Note, that initially the fidelity grows and, as expected, numerical
simulations show that the metastable region increases as the
perturbation of the angles is decreased. Numerical evidence suggests
chaotic dynamics also in the two-qubit case, but a rigorous proof is
still lacking.

\section{Conclusions}\label{concl}

Complex chaos realized by measurement conditioned iterative
deterministic quantum maps takes place in an ensemble, where the
ensemble size decreases exponentially with time if we started with a
finite number of copies. This is due to the fact that the measured
part of the system cannot take part in the dynamics anymore. In
purification protocols only the first few steps are used. A possible
application of high sensitivity to small differences in the initial
states is to apply it as a Schr\"odinger microscope and discriminate
between to close state.

There are several open questions about the conditions leading to
complex chaos. When we include mixed states, the parameter space
further increases.  The proof of chaos is missing here. One could
deviate from von Neumann measurements by measuring a larger part of
the ensemble jointly and use the gained information either for some
form of feedback, as proposed in \cite{Lloyd2000} or as a selection
criterium, like in the presented scheme.

\section*{Acknowledgments}
Support by the Czech and Hungarian Ministries of Education
(CZ-2/2005), by GA\v CR 202/04/2101, by MSMT LC 06001, by the European
Union, by the Hungarian Scientific Research Fund (T043287 and T049234)
and by DAAD is acknowledged.

\vfill\eject

\end{document}